\documentclass[11pt]{scrartcl}
\usepackage{geometry}                
\usepackage{newpxtext} \usepackage[euler-digits]{eulervm}

\usepackage{graphicx}
\usepackage{amssymb}
\usepackage{epstopdf}
\usepackage{amsmath}
\usepackage{feynmp}

\usepackage{dsfont} 
\usepackage{verbatim}
\usepackage{mathrsfs} 
\usepackage{amsthm}
\usepackage{amscd}
\usepackage{color, soul}

\usepackage{tikz}
\usetikzlibrary{decorations.markings}

\newcommand{\be}{\begin{equation}}
\newcommand{\ee}{\end{equation}}
\newcommand{\eps}{\epsilon}
\newcommand{\bal}{\begin{aligned}}
\newcommand{\eal}{\end{aligned}}

\usepackage{hyperref}
\hypersetup{
    colorlinks=true,
    linkcolor=blue,
    filecolor=black,      
    urlcolor=blue,
    citecolor=magenta,
}

\newtheorem*{Prop*}{Proposition}

\theoremstyle{definition} 
\theoremstyle{definition} 
\theoremstyle{definition} 
\theoremstyle{definition} \newtheorem*{Rem*}{Remark}

\DeclareMathOperator{\Si}{Si}

\numberwithin{equation}{section}

\title{Convergent and Divergent Series in Physics}
\subtitle{Lectures of the 22nd ``Saalburg'' Summer School (2016)}
\author{A short course by Carl Bender${}^\dagger$\\{Notes written and edited by Carlo Heissenberg${}^\ast$}\vspace{-50pt}}

\date{}
\begin{document}
\maketitle

{\small	\noindent${}^\dagger$ Department of Physics, Washington University, St. Louis, Missouri 63130, USA.
	\href{mailto:cmb@wustl.edu}{cmb@wustl.edu}
	
	\vspace{5pt}
\noindent	${}^\ast$ Scuola Normale Superiore, Piazza dei Cavalieri, 7 56126 Pisa, Italy.
	 \href{mailto:carlo.heissenberg@sns.it}{carlo.heissenberg@sns.it}}

\section*{Introduction}
The aim of this short series of lectures is to cover a few selected topics in the theory of perturbation series and their summation. 
The issue of extracting sensible information from a formal power series arises quite naturally in physics: it is often impossible, when tackling a given (hard) problem, to obtain an exact answer to it and it is typically useful to introduce some small parameter $\epsilon$ in order to obtain successive approximations to the correct solution order by order in $\epsilon$, starting from an exact known solution to the problem with $\eps=0$. The result will be therefore given by the formal expression
\vspace{-5pt}$$
\sum_{n=0}^\infty a_n \epsilon^n, 
\vspace{-5pt}$$
and one hopes to be able to give a meaning to such a series as $\epsilon$ approaches $1$, in order to recover the correct answer to the original problem. This last step of the program is especially difficult to carry out in many cases and one  needs some nontrivial strategy in order to extract relevant physical information from a perturbative series.

The first part of these lectures will be devoted to the description of strategies for accelerating the rate of convergence of \emph{convergent} series, namely Richardson's extrapolation,
and the Shanks transformations, and will also cover a few interesting techniques for employing the Fourier series in a convenient way.

In the second part we will essentially focus on \emph{divergent} series, and on the tools allowing one to retrieve information from them, despite the fact that their sum is obviously ill-defined. These techniques include Euler summation, Borel summation and generic summation; however, as we shall discuss, the most powerful tool for summing a divergent series is the method of continued functions, and in particular the Pad\'e theory based on continued fractions.

\tableofcontents

\section{A few examples}
As a first example of a nontrivial problem, whose solution can be approximated with asymptotic techniques, consider the following equation:
\be
x^5+x-1=0.
\ee
Even though it is clear that this equation indeed has one real root, since $p(x) = x^5 +x-1$ goes from $-\infty$ to $+\infty$ with positive derivative, such a solution is not known in any exact closed form; one can nevertheless try to approximate it in a perturbative way by introducing a ``small'' parameter $\eps$ in the following way
\be\label{eq: x^5eps}
x^5+\eps x-1=0.
\ee
The unperturbed problem, defined by $\eps=0$, is exactly solvable: the corresponding real root is $x=1$. Furthermore, one can try to obtain the solution as a formal power series in $\epsilon$
\be\label{eq: serie1}
x(\epsilon) = \sum_{n=0}^\infty a_n x^n,
\ee
where $a_0=1$. Substituting this into \eqref{eq: x^5eps} and setting to zero the coefficient of $\epsilon^n$, order by order in $\epsilon$, yields the equations
\be
\begin{aligned}
5a_1+1&=0,\\
5a_2 + 10 a_1^2 + a_1&=0,\\
\end{aligned}
\ee
and so on, which can be solved recursively in order to obtain all the $a_n$. It is also possible to obtain a closed form for them and to prove that the convergence radius of the series \eqref{eq: serie1} is 
\be
\rho = \left(\frac{5}{4}\right)^{4/5}>1.
\ee
This shows that in principle the solution can be  approximated with arbitrary accuracy by computing more and more terms of the series after setting $\epsilon=1$.

There is however another way of inserting a small parameter $\eps$ into this problem, such that the corresponding unperturbed problem $\eps=0$ is exactly solvable, namely
\be
\epsilon x^5 + x - 1 =0.
\ee
However, the perturbative series for the solution $x(\eps)$, in this case, only converges with radius 
\be
\rho=\frac{4^4}{5^5}<1
\ee
and hence the naive idea of just adding up the terms of such a series for $\epsilon=1$ has no hope of succeeding, since the series \emph{diverges} for $\epsilon=1$. As we shall see, divergent series also quite often arise when tackling problems in physics, and their occurrence can be motivated by physical considerations.

As an exercise, we can ask ourselves what would happen if we considered the problem of finding the \emph{complex} roots of the above polynomial of fifth degree:
\be
z^5+z-1=0.
\ee 
We can again insert $\eps$ in two ways:
\be
z^5+\eps z-1=0
\ee 
corresponds for instance to the unperturbed problem
\be
z^5 -1=0,
\ee
solved by $z=1, e^{\pm i2\pi/5}, e^{\pm i4\pi/5}$, and higher orders in $\eps$ will give us small corrections around these five values; we can however consider
\be\label{eq: complex_roots}
\eps z^5 + z -1=0,
\ee 
which gives, for $\eps=0$,
\be\label{ovvio}
z=1.
\ee
The second kind of perturbative approach is called \emph{singular} perturbation since the problem changes abruptly when we turn off the perturbation parameter: in this case, the polynomial changes its degree and 4 of its 5 complex roots seem to have disappeared. This is analogous to considering the semiclassical limit of quantum mechanics, where the Schr\"odinger equation
\be
-\frac{\hbar}{2m}\nabla^2 \psi(x) + V(x) \psi(x) = E\psi(x)
\ee 
becomes an algebraic equation as $\hbar \to0$. 

To see what happens to the complex roots of equation \eqref{eq: complex_roots}, we can make the following observation. As $\eps\to0$, there are only two allowed behavior of $z$ which can give rise to cancellation between the terms of the equation: the first is ``$z\sim$constant'', which leads us to the result \eqref{ovvio}, but it could also happen that $z$ becomes very large, and then the $-1$ term becomes negligible. In the latter case,
\be
\eps z^5 +z=0 \implies z\sim \eps^{-1/4}
\ee
which indicates that, as $\eps\to\infty$, the complex roots of the equation run off to infinity as $\eps^{-1/4}$.
Furthermore, letting now $z(\eps) = \eps^{-1/4}s(\eps)$, we can reduce this singular perturbation problem to a non-singular one
\be
s^5 + s - \eps^{1/4}=0
\ee
where $\delta=\eps^{1/4}$ is the natural perturbation parameter.
\newpage 
\section{Summing Convergent Series Efficiently}
In this section, we shall describe a few tools and techniques for dealing with convergent series in a convenient way.

\subsection{Shanks transformation}
Consider the series
\be
\sum_{n=0}^\infty a_n.
\ee
Such a series is said to be convergent if its partial sums
\be
A_N = \sum_{n=0}^N a_n
\ee
have a well-defined and finite limit $s$, called the sum of the series:
\be
\lim_{N\to\infty} A_N = s < \infty.
\ee
A series is said to diverge whenever it does not converge.
Convergence can be absolute, \emph{i.e.} in absolute value 
\be
\sum_{n=0}^\infty |a_n|= s < \infty, 
\ee
or only conditional, that is when, although the series series converges, the corresponding series of absolute values diverges. A typical case of conditional convergence provided by
\be\label{eq: Taylor_log2}
\sum_{n=1}^\infty (-1)^{n+1}\frac{1}{n} = \log 2,
\ee
which is alternating in sign and oscillates around its sum. 

To approximate the sum of the series, one can of course add up more and more of its terms; however, this is not the best idea we can come up with!
A numerically much more efficient strategy allowing one to extract information from an \emph{oscillating} series is the following: since we know that the series displays smaller and smaller oscillations around its sum, it is reasonable to assume that the partial sums approximate the actual sum $s$, for example, with the following asymptotic behavior
\be
A_n\sim s+ k r^N, \quad \text{as }N\to\infty,
\ee
where $|r|<1$ (in fact, $r<0$ since the series oscillates). In turn, this suggests a way of approximating $s$ itself:
\be
\bal
A_{N-1} &= s + kr^{N-1},\\
A_N &= s + kr^N,\\
A_{N+1} &= s + kr^{N+1}.
\eal
\ee
Thus
\be
\frac{A_N-s}{A_{N-1}-s} = R = \frac{A_{N+1}-s}{A_N-s}
\ee
and finally 
\be
s = \frac{A_{N+1}A_{N-1}-A_N^2}{A_{N+1}-2A_N+A_{N-1}}.
\ee
This is the \textbf{Shanks transformation}. It is remarkable that, although the series \eqref{eq: Taylor_log2} takes quite lot to give a reasonable approximation to the correct answer, the Shanks transformation already provides a satisfactory result within the first few partial sums.

The intuitive idea behind this improvement in the rate of convergence is that, if we think of $\eps$ as  a complex number, performing the Shanks transformation moves singularities further away from the origin in the complex plane.

\subsection{Richardson extrapolation}
What if the (convergent) series we are trying to sum is \emph{not} oscillating? An interesting case of a converging series with all positive terms is given, for example, by the Riemann zeta function $\zeta(s)$ computed for $s=2$,
\be
\sum_{n=1}^\infty\frac{1}{n^2} = \frac{\pi^2}{6}.
\ee
Is there some strategy we can employ in order to improve the convergence of such a series as well?
Let us try to isolate the error term for the $N$th partial sum: for large $N$ we may use the following approximation
\be
\bal
s = \sum_{n=0}^\infty \frac{1}{n^2} &= A_N + \sum_{n=N+1}^\infty \frac{1}{n^2}\\
&\sim A_N + \int_N^\infty \frac{dn}{n^2}\\
&\sim A_N + \mathcal O\left( \frac{1}{N} \right).
\eal
\ee
This suggests that, for a generic series with positive terms, we could write
\be\label{eq: Richardson_appr}
A_N \sim s + \frac{a}{N}+ \frac{b}{N^2} + \frac{c}{N^3} + \ldots
\ee
and truncate this expansion after a few leading terms. Choosing to keep only $a/N$, and writing down the approximation for $N$ and $N+1$, leads to
\be\bal
A_N &= s + \frac{a}{N},\\
A_{N+1} &= s + \frac{a}{N+1}.
\eal\ee
Thus
\be\bal
N A_N  &= N s + a,\\
(N+1) A_{N+1} &= (N+1) s + a, 
\eal\ee
and we finally obtain, subtracting the first equation from the second one,
\be
s = (N+1) A_{N+1} - N A_N.
\ee
This is called the \textbf{first Richardson extrapolation}, and again one can see by performing simple numerical experiments that its properties of convergence are much better compared direct addition of the $a_n$.

One can in fact keep more terms in the expansion \eqref{eq: Richardson_appr}, and obtain an even faster convergence towards the sum of the series:
\be\bal 
A_N &= s + \frac{a}{N} + \frac{b}{N^2},\\
A_{N+1} &= s + \frac{a}{N+1} + \frac{b}{(N+1)^2},\\
A_{N+2} &= s + \frac{a}{N+2} + \frac{b}{(N+2)^2}.
\eal\ee
Hence
\be\bal 
N^2 A_N &= N^2 s + N a + b,\\
(N+1)^2 A_{N+1} &= (N+1)^2 s + (N+1) a + b,\\
(N+2)^2 A_{N+2} &= (N+2)^2 s + (N+2) a + b,
\eal\ee
and finally, summing the first with the last equations and subtracting twice the second equation, we get
\be
s=\frac{1}{2}\left[ (N+2)^2 A_{N+2} - 2(N+1)^2 A_{N+1} + N^2 A_N \right]
\ee
which is the second Richardson extrapolation. One can in principle go on with this procedure obtaining
\be
s = \frac{1}{3!}\left[ (N+3)^3 A_{N+3} - 3(N+2)^3 A_{N+2} + 3(N+1)^3 A_{N+1} - N^3 A_N \right]
\ee
or for $k\in \mathbb N$
\be
s = \frac{1}{k!}\sum_{l=0}^k (-1)^{k-l} (N+l)^k  \binom{N}{l} A_{N+l}.
\ee
We shall not deal with the rigorous results concerning this improved summation methods, but it should be noticed that their efficacy is well-established in most physically relevant applications; furthermore, let us mention that the celebrated Simpson rule and Romberg's integration can be both seen as applications of the Richardson extrapolation. 
\newpage
\section{Fourier Series}
Although the importance of Fourier series in physically relevant problems is well-known, the power of such a tool is not always fully appreciated in elementary courses. The purpose of this section is to provide an overview of techniques allowing to exploit the Fourier series to its full potential when tackling physical problems; in particular, we shall see how the Gibbs phenomenon, which is often regarded as a pathological feature of the Fourier series, can be in fact used in order to solve Cauchy problems with nonstandard boundary conditions.

\subsection{Heat equation and Fourier series}\label{sub: Fourier_naive}
Consider the Cauchy problem corresponding to the one-dimensional diffusion equation, for $x\in[0,\pi]$ and $t>0$,
\be\label{eq: Cauchy_Fourier}
u_t(x,t)= u_{xx}(x,t),
\ee
with initial and boundary data given by 
\be\begin{aligned}
u(x,0) &= f(x),\\
u(0,t) &= g(t),\\
u(\pi,t) &= h(t).
\end{aligned}
\ee 
A first idea to tackle such a problem could be to expand $u(x,t)$ as a power series in $t$
\be\bal
u(x,t) &= \sum_{n=0}^\infty a_n(x) t^n,\\
a_0(x) &= f(x),
\eal\ee
but on physical grounds this approach is highly unlikely to give a sensible result: a Taylor series can only converge in a symmetric interval, and this means that, if the series were convergent, we would be able to diffuse backwards in time (that is, we could reconstruct the initial conditions that would give the result in the present time under the diffusion process). This is nonsensical since diffusion is an \emph{irreversible} process.

A better idea is to proceed as follows. Let us assume for now that $g(t) = h(t) = 0$, \emph{i.e.} homogeneous boundary conditions in time, and let us look for a separated-variable solution of the type $u(x,t) = A(x)B(t)$. Substituting this into $u_t = u_{xx}$ gives
\be
\frac{B'(t)}{B(t)} = \frac{A''(x)}{A(x)},
\ee
but since the left-hand and the right-hand side  depend on $x$ and $y$ separately, this can only be satisfied when
\be
B'(t) = -c B(t),\qquad A''(x) + c A(x)=0
\ee
for some constant $c$, giving thus the general solutions
\be
B(t) = e^{-ct},\qquad A(x) = a \sin(\sqrt{c}x) + b \cos(\sqrt{c}x),
\ee
where the unphysical case $c<0$, which would let $B(t)$ diverge as $t\to+\infty$, has been excluded;
imposing now the boundary conditions $u(0,t) = 0 = u(\pi, t)$ selects the values $\sqrt c = n$, for $n\in \mathbb N$. Hence
\be
B_n(t) = e^{-n^2 t}, \qquad A_n(x) = a_n \sin(nx).
\ee
Now, by linearity of the equation, one is led to consider the following general form of the solution
to the homogeneous problem, subject to $u(x,0)= f(x)$,
\begin{align}
u(x,t) &= \sum_{n=1}^\infty a_n e^{-n^2t} \sin(nx),\label{eq: naive}\\
f(x) &= \sum_{n=1}^\infty a_n \sin(nx), \label{eq: fFourier}
\end{align}
which is indeed the Fourier (sine) series of $f(x)$.
One is left with the task of computing $a_n$, and this is usually done as follows: since $\{ \sin (nx),\ n\in\mathbb N \}$ is an orthogonal set of functions over $[0, \pi]$,
\be
\int_0^\pi \sin(nx) \sin(mx)\, dx = \frac{\pi}{2}\,\delta_{nm},
\ee
then, multiplying \eqref{eq: fFourier} by $\sin (mx)$ and integrating over $x$, one has
\be
a_n = \frac{2}{\pi} \int_0^\pi f(x) \sin(nx)\, dx.
\ee

However, this strategy runs into difficulties if we are confronted with functions without sufficient regularity conditions ensuring that we can indeed exchange orders of summation and integrations in the last step! 

One can prove that for functions $f\in C^1([0,\pi])$ which vanish at the endpoints $f(0)=0=f(\pi)$, this operation is well-defined and the Fourier series converges uniformly to $f(x)$ on $[0,\pi]$. For instance, this is the case for the function $f(x) = x(\pi-x)$.

On the other hand, already for the constant function $f(x) = 1$ on $[0,\pi]$, this does not hold anymore, and we need to find a better strategy: near the endpoints, oscillations in the partial sums of the Fourier series occur which grow bigger and bigger as one approaches $0$ and $\pi$. This is known as the \emph{Gibbs phenomenon}.

\subsection{Convergence of a Fourier series}
It is convenient, instead, to adopt a somewhat more constructive approach: let us simply \emph{define} $a_n$ as 
\be\label{eq: a_n(def)}
a_n = \frac{2}{\pi} \int_0^\pi f(x) \sin(nx) dx
\ee
for any continuous $f$ on the interval $[0,\pi]$ (which need not vanish at the endpoints). This integral clearly exists and we then inquire as to whether the corresponding Fourier series 
\be
\sum_{n=1}^\infty a_n \sin(nx) 
\ee
is convergent or not. Indeed, such a series
converges to $f(x)$
at each point $x\in (0,\pi).$ Consider for instance the constant function $f(x)=1$ on $[0,\pi]$. Then the corresponding Fourier coefficients are
\be
a_n = \frac{2}{\pi} \int_0^\pi \sin(nx) dx = \begin{cases}
0 &\text{for even }n,\\
\dfrac{4}{n\pi} &\text{for odd }n.
\end{cases}
\ee
The Fourier series then reads
\be
\sum_{l=0}^\infty \frac{4 \sin\left[(2l+1)x\right]}{\pi(2l+1)}.
\ee
Denoting by $A_N(x)$ the $N$th partial sum, we can compute
\be\bal
A'_N(x) &= \frac{4}{\pi} \sum_{l=0}^{N} \cos\left[(2l+1)x\right] =\frac{2}{\pi}e^{-(2N+1)x}\sum_{l=0}^N \left(e^{i2x}\right)^l\\
&= \frac{2}{\pi}e^{-(2N+1)x}\, \frac{1-e^{i2(2N+2)x}}{1-e^{i2x}} = \frac{2\sin\left[2(N+1)x\right]}{\pi\sin x},
\eal\ee
and hence
\be
A_N(x) = \frac{2}{\pi} \int_0^x \frac{\sin\left[2(N+1)s\right]}{\sin s} ds.
\ee
Let us rewrite this as
\be
A_N(x) = \frac{2}{\pi} \int_0^x \sin\left[2(N+1)s\right]\left( \frac{1}{\sin s} - \frac{1}{s}\right) ds + \frac{2}{\pi} \int_0^x \frac{\sin\left[2(N+1)s\right]}{s} ds;
\ee
the first term now vanishes as $N\to\infty$ due to the Riemann-Lebesgue lemma, since $1/\sin s - 1/s $ is continuous on $[0,x]$, whereas the second term gives\footnote{
If instead one takes the \emph{double scaling} limit $N\to\infty$ and $x\to0$ with $2(N+1)x=\alpha$ for fixed $\alpha$ in \eqref{eq: double_scaling}, one sees that the series approaches the entire analytic function 
$$
\frac{2}{\pi} \int_0^\alpha \frac{\sin s}{s}ds = \frac{2}{\pi} \Si (\alpha).
$$
}
\be\label{eq: double_scaling}
\frac{2}{\pi} \int_0^{2(N+1)x} \frac{\sin s}{s} ds \xrightarrow[N\to\infty]{} \frac{2}{\pi}\int_0^\infty \frac{\sin s}{s}ds = 1.
\ee
Such a proof can be in fact extended to any continuous $f$ over $[0,\pi]$ by splitting the interval $[0,\pi]$ into $N$ sub-intervals: first, one applies this result to simple functions of the form $\varphi(x) = \sum c_n \chi_n(x)$, where $\chi_n(x)$ is the characteristic function of the $n$th sub-interval of $[0,\pi]$, and then one recalls that continuous functions can be always approximated by simple functions with arbitrary accuracy.

Do we have any information concerning the rate convergence of the Fourier series? Integrating a few times by parts the definition \eqref{eq: a_n(def)} of $a_n$ we obtain:
\be\bal
a_n &= \frac{2}{\pi}\left[-\frac{1}{n}f(x)\cos(nx)\Big|_0^\pi + \frac{1}{n}\int_0^\pi f'(x) \cos(nx) dx \right]\\
&= \frac{2}{\pi}\left[\frac{c^{(n)}_1}{n} + \frac{c^{(n)}_2}{n^3} + \ldots \right],
\eal\ee
where
\be
c^{(n)}_1 = (-1)^{n+1}f(\pi)+f(0),
\ee
$c^{(n)}_2$ is a similar coefficient involving $f''(x)$ computed at the endpoints, and further subleading contributions in $1/n$ have been omitted. This argument allows us to estimate the asymptotic behavior of the series coefficients: If the function $f(x)$ does not vanish at the endpoints, then 
\be
a_n\sim \frac{2 c_1^{n}}{\pi n},
\ee
indicating slow, in fact only conditional, convergence,
whereas if it does, the coefficients behave as
\be
a_n\sim \frac{2 c_2^{n}}{\pi n^3},
\ee
which shows that the series converges absolutely and rapidly. Notice that the converse is also true: If the Fourier series converges rapidly, then the function must vanish at the endpoints.

This suggests a way of accelerating the convergence of a given Fourier series. Consider for instance the Fourier coefficients
\be
a_n = e^{1/n}-1 \sim \frac{1}{n}, \quad\text{as    }n\to\infty. 
\ee 
The Fourier series corresponding to $a_n$ is therefore slowly convergent and we are in presence of the Gibbs phenomenon. To overcome this obstacle, we exploit the explicit knowledge of the following Fourier series:
\be
g(x) = \frac{\pi-x}{2} = \sum_{n=1}^\infty \frac{\sin(nx)}{n}; 
\ee
adding and subtracting $g(x)$ then yields
\be
f(x) = g(x) + \sum_{n=1}^\infty \left(a_n - \frac{1}{n}\right)\sin(nx).
\ee
Now, the series on the right-hand side has coefficients $a_n-1/n$ displaying \emph{fast} convergence; furthermore, as a consequence, it must give vanishing contributions at the endpoints, hence
\be
f(0) = g(0) = \frac{\pi}{2},\quad f(\pi) =g(\pi)=0.
\ee

\subsection{Exploiting the Gibbs phenomenon}
Let us go back the original problem of solving equation \eqref{eq: Cauchy_Fourier}. Another tempting possibility could be to expand $u(x,t)$ as the following Fourier series
\be
u(x,t) = \sum_{n=1}^\infty a_n(t) \sin(nx),
\ee 
which, as we saw, converges pointwise for each $x$ in the interior of the desired interval. Then carelessly substituting in the heat equation yields
\be\label{eq: not_convergent}
\sum_{n=1}^\infty \dot a_n(t) \sin(nx) = - \sum_{n=1}^\infty a_n(t) n^2 \sin(nx)
\ee
hence
\be
\dot a_n(t) = -n^2 a_n(t) \implies a_n(t) = a_n(0) e^{-n^2t},
\ee
which gives again \eqref{eq: naive} and \eqref{eq: fFourier}.

Note however that the right-hand side of \eqref{eq: not_convergent} is ill-defined precisely when the boundary conditions $g(t)$ and $h(t)$ are nontrivial, since then the $a_n(t)$ will be slowly converging and $n^2a_n(t)$ will give rise do a \emph{divergent} Fourier series. Therefore, this approach cannot give any new answer with respect to the one we employed in Section \ref{sub: Fourier_naive}.

More generally, what we can do is again differentiate the definition of $a_n(t)$:
\be\bal
a_n(t) &= \frac{2}{\pi} \int_0^\pi \sin(nx) u(x,t) dx,\\
\dot a_n(t) &= \frac{2}{\pi} \int_0^\pi \sin(nx) u_t(x,t) dx\\
&= \frac{2}{\pi} \int_0^\pi \sin(nx) u_{xx}(x,t) dx,
\eal\ee
where the heat equation has been used in the last step. Notice that the derivative can be legitimately pulled under the integral sign, since this integral converges absolutely. Now we integrate by parts twice:
\be\bal
\dot a_n(t) &= \frac{2}{\pi} \left[ \sin(nx) u_x(x,t)\big|_0^\pi - n \int_0^\pi \cos(nx) u_x(x,t)dx\right]\\
&= \frac{2}{\pi} \left[ -n \cos(nx) u(x,t)\big|_0^\pi - n^2 \int_0^\pi \sin(nx) u(x,t) dx\right]\\
&= \frac{2n}{\pi} \left[(-1)^{n+1} h(t) + g(t)\right] - n^2 a_n(t). 
\eal\ee
Notice how in this way the boundary conditions have now consistently appeared: this is a consequence of the Gibbs phenomenon and it allows us to express in Fourier series even solutions which do \emph{not} vanish at the endpoints of the interval! Defining for simplicity $\phi(t) \equiv (-1)^{n+1} h(t)+ g(t)$, we need to solve
\be
\dot a_n(t) + n^2 a_n(t) = \frac{2n}{\pi}\phi(t),
\ee
\emph{i.e.}
\be
\frac{d}{dt}\left[a_n(t) e^{n^2 t}\right]= \frac{2n}{\pi}e^{n^2t}\phi(t)
\ee
and finally
\be\label{eq: Fourier_Gibbs}
a_n(t) = e^{-n^2 t}\left[a_n(0) + \frac{2n}{\pi} \int_0^t e^{n^2s}\phi(s) ds \right],
\ee 
where $a_n(0)$ is given by the Fourier expansion of $f(x)$. 

Now, we know that this series converges for $x\in(0,\pi)$, and we also know that such convergence cannot be but conditional, with $a_n(t)$ falling off like $1/n$ as $n\to\infty$. This can be indeed checked explicitly, again integrating by parts: considering only the second term in \eqref{eq: Fourier_Gibbs} since the first one is exponentially suppressed as $n\to\infty$, we have
\be\bal
\frac{2n}{\pi} e^{-n^2 t}\int_0^t e^{n^2s}\phi(s) ds &= \frac{2n}{\pi} e^{-n^2 t}\left[\frac{1}{n^2}\phi(s)e^{n^2 s}\Big|_0^t - \frac{1}{n^2}\int_0^t e^{n^2s} \dot \phi(s) ds\right]\\
&= \frac{2}{\pi n} \phi(t) +\ldots
\eal\ee
where further subleading contributions in $1/n$ have been omitted.
Thus, 
\be
a_n(t) \sim \frac{2}{\pi n} \phi(t),\quad \text{as }n\to\infty.
\ee
Since we have extimated the asymptotic behavior of the Fourier coefficients, we may exploit this knowledge to accelerate the convergence of the series, as we have done before: consider
\be
a_n(t) = \frac{2}{\pi n} \phi(t) + b_n(t);
\ee
now, the first term has isolated the $1/n$ contribution to the series, whereas the $b_n(t)$ are rapidly converging as $1/n^3$, for $n\to\infty$. As a matter of fact, we also know how to sum the series corresponding to the first term,
\be
\frac{2}{\pi}\sum_{n=1}^\infty \frac{\sin(nx)}{n} \phi(t) = \frac{\pi-x}{2}\phi(t). 
\ee
Note also that this procedure can be repeated in principle as many times as one likes by integrating by parts and using Fourier series identities to improve the convergence at each step.
\newpage
\section{Divergent Series}
Let us now turn the study of divergent series. Is there any meaning we can assign to this kind of series? The answer of course is highly non-unique: in principle we could just define any divergent series to be equal to our favourite number. We shall see however that one can motivate certain types of formal summation by requiring suitable underlying mathematical properties. In fact, it will turn out that summing divergent series is often much easier than summing the convergent ones!

Examples of divergent series are given for instance by the following ones:
\be\bal\label{eq: examples_div}
&1-1+1-1+\ldots\ ,\\
&1+1+1+1+\ldots\ ,\\
&1+0-1+1+0-1+\ldots\ ,\\
&1+2+4+8+\ldots\ ,\\
&1!-2!+3!-4!+5!-\ldots\ \ .
\eal\ee
A first observation while dealing with divergent series, and in fact even with conditionally convergent ones, is that the usual properties of arithmetic do not admit a simple extension to operations with an \emph{infinite} number of terms: For instance, carelessly enforcing commutativity on the series
\be
\log 2= \sum_{n=1}^\infty (-1)^{n+1}\frac{1}{n}
\ee
allows us to obtain any chosen real number. Indeed, let us first rewrite the series as the sum of its alternating positive and negative terms
\be
\sum_{n=1}^\infty (-1)^{n+1}\frac{1}{n} = p_1 + n_1 + p_2 + n_2 +\ldots
\ee
and take for instance the number $12$. We can add up the $p_1+p_2+\ldots$ until the partial sum becomes greater than $12$; then start adding $n_1+n_2+\ldots$ until the partial sum gets below $12$. Repeating this procedure, the partial sums will oscillate around $12$, with smaller and smaller amplitude, indicating that the reordered series converges to $12$. 
Similarly, also associativity cannot be effortlessly extended to operations among series: we shall see that, for example, the first and third series in \eqref{eq: examples_div} give different results.

Finally, we should remark that there is nothing refraining a divergent series from having \emph{infinite} sum, a notable example being the harmonic series:
\be
\sum_{n=1}^\infty  \frac{1}{n} = \infty.
\ee
In fact, there are even physical situations in which the correct answer is infinity.

 Take for instance the problem of piling up bricks of equal length $l=2$ in order to build a staircase going as high as possible and with the greatest possible overhang. 
Take the first brick, and place under it a second one in such a way that the first one has an overhang of $1$: in this way, the center of mass falls at $3/2$ from the tip of the lower brick, and the system is stable since this value is lower than the brick's length ($l=2$).
Now proceed by placing a third brick under the first two, so that the second brick has an additional overhang of $1/2$ with respect to the new brick: now the center of mass lies at  $5/3$. 
Poceeding this way, we see that at the $n$th step the total overhang will be
\be
1+\frac{1}{2}+\frac{1}{3}+\ldots+\frac{1}{n},
\ee
and the position of the center of mass will be $(2n+1)/(n+1)$, still lower than $2$.

\subsection{Euler summation}
Consider the divergent series
\be
\sum_{n=0}^{\infty} a_n.
\ee
The procedure suggested by Euler, in order to assign a meaning to it, is to replace at first this series by the corresponding power series
\be
\sum_{n=0}^{\infty} a_n x^n.
\ee
If such series is convergent for $|x|<1$ and if its limit $x\to1^-$ exists, then one defines the \textbf{Euler summation} of the original series as
\be
E\left(\sum_{n=0}^\infty a_n\right) = \lim_{x\to1^-} \sum_{n=0}^\infty a_n x^n.
\ee 
Applying for instance this method to the divergent series
\be\label{eq: +-1}
\sum_{n=0}^\infty (-1)^n = 1-1+1-1+\ldots
\ee
one finds 
\be
E\left( \sum_{n=0}^\infty (-1)^n \right) = \lim_{x\to1^-}\sum_{n=0}^{\infty}(-x)^n = \lim_{x\to1^-}\frac{1}{1+x} = \frac{1}{2}.
\ee
In this case, such a procedure can be also thought of as taking an average: since the $N$th partial sum $A_N$ is $1$ if $N$ is even and $0$ if $N$ is odd, one has
\be
\lim_{K\to\infty} \frac{1}{K+1}\sum_{N=0}^K A_N = \frac{1}{2}.
\ee
The more general case of the series
\be
\sum_{n=1}^\infty (-1)^{n-1}n^p = 1-2^p+3^p-4^p+\ldots\ ,
\ee
where $p$ is a positive integer, can be tackled as follows (absolute convergence for $|t|<1$ allows us to interchange summation and differentiation): 
\be
\sum_{n=1}^\infty (-1)^{n-1}n^p t^{n} =\left( t \frac{d}{dt} \right)^p  \sum_{n=1}^\infty (-1)^{n-1} t^{n}= \left( t \frac{d}{dt} \right)^p \frac{t}{1+t},
\ee
so that, letting $t=e^x$,
\be\label{compare!}
E\left( \sum_{n=1}^\infty (-1)^n n^p\right) = \frac{d^p}{dx^p}\frac{1}{1+e^{-x}}\bigg|_{x=1}\!\!=s_p.
\ee
This gives the sequence
\be
\{s_p;\ p=0,1,\ldots\}=
\left\{\frac{1}{2}\, ,\ 
\frac{1}{4}\, ,\
0\, ,\ 
-\frac{1}{8}\, ,\ 
0\, ,\
\frac{1}{4}\, ,\
0\, ,\ 
-\frac{17}{16}\, ,\
0\, ,\
\frac{31}{4}\, ,\
0\, ,\
-\frac{691}{8}\, ,\ \ldots \right\},
\ee
which is closely related to the Bernoulli numbers.

\subsection{Borel summation}
Borel's proposal for regularizing a given divergent series, instead, takes inspiration from the integral formula 
\be
n! = \int_0^\infty t^n e^{-t} dt.
\ee 
Dividing and multiplying by $n!$, and formally interchanging the order of summation and integration, one has
\be
\sum_{n=0}^\infty a_n = \sum_{n=0}^\infty \int_0^\infty \frac{a_n}{n!} t^n e^{-t}dt
\longmapsto
 \int_0^{\infty} \left( \sum_{n=0}^\infty \frac{a_n}{n!}t^n\right) e^{-t}dt.
\ee
We see that the series appearing now inside the integral has now much better chances of converging, at least for small $t$. Indeed, one can generalize Borel's approach by substituting the formulas for, say,  $(2n)!$ or $(6n)!$ which improve convergence even more. This motivates the following definition of \textbf{Borel summation}:
\be
B\left( \sum_{n=0}^{\infty} a_n \right) = \int_0^{+\infty} g(t) e^{-t} dt, \qquad \text{where}\qquad g(t) = \sum_{n=0}^\infty \frac{a_n}{n!}t^n; 
\ee
here, it is assumed that the series defining $g(t)$ converges at least in some open interval on the positive real axis, and that the integral $\int_0^\infty g(t)e^{-t}dt$ is finite.

As an example, consider again the series \eqref{eq: +-1}: the function $g(t)$ is given by
\be
g(t) = \sum_{n=0}^\infty \frac{(-t)^n}{n!} = e^{-t},
\ee
and hence
\be
B\left( \sum_{n=0}^{+\infty} (-1)^n \right) = \int_0^\infty e^{-2t}dt = \frac{1}{2}.
\ee
As a further example, let us consider
\be
\sum_{n=1}^\infty (-1)^{n-1} n = 1-2+3-4+\ldots\ \ .
\ee
The corresponding $g(t)$ is
\be
g(t)= \sum_{n=1}^\infty \frac{(-1)^{n-1}}{(n-1)!}t^n=t e^{-t},
\ee
and therefore 
\be
B\left( \sum_{n=1}^{+\infty} (-1)^{n-1} n\right) = \int_0^\infty te^{-2t}dt = \frac{1}{4}.
\ee
What about the series
\be
\sum_{n=1}^\infty (-1)^{n-1}n^2=1-2^2+3^2-4^2+\ldots\ ?
\ee
In this case, interchanging summation and differentiation by uniform convergence, we get
\be\begin{aligned}
g(t)=\sum_{n=1}^{\infty}\frac{(-1)^{n-1}}{(n-1)!}nt^n
= t \frac{d}{dt} \sum_{n=1}^\infty \frac{(-1)^{n-1}}{(n-1)!}t^n
= t \frac{d}{dt}\left(t e^{-t}\right).
\end{aligned}\ee
Correspondingly, 
\be\begin{aligned}
B\left( \sum_{n=1}^{+\infty} (-1)^{n-1} n^2\right)=\int_0^\infty te^{-t} \frac{d}{dt}(te^{-t})dt
=\frac{1}{2}\int_0^{\infty} \frac{d}{dt}(t^2 e^{-2t})dt
=0.
\end{aligned}\ee
Thus, the Borel sum of $1-2^2+3^2-4^2+\ldots$ vanishes identically, in agreement with \eqref{compare!}, which is a rather amusing result.
The generalization of these formulas to a given integer power $p>0$ is given by
\be\label{compare!!}\
B\left(\sum_{n=1}^\infty (-1)^{n-1}n^p\right) = \int_0^\infty t e^{-t}\left(t\frac{d}{dt}\right)^{p-1}\!\!\!\left(te^{-t}\right) \frac{dt}{t}=s_p.
\ee
We see therefore that the two different methods proposed by Euler and Borel in fact agree, at least on these rather simple examples, and we can wonder if there is any underlying, unifying mathematical framework linking the two procedures. This will be the subject of the next section.

\subsection{Generic summation}
Now we will try to define a generic class of summation techniques by treating them as ``black boxes'' and only imposing generic properties on their behavior. For this purpose, consider a given summation method $S$ and \emph{assume} it does give some finite answer for the series you are interested in. First, we want that, taking out a finite number of terms of the series, the final result be consistent with such a subtraction: 
\begin{itemize}
\item[(1)] (\textit{Finite additivity})
\be
S\left(\sum_{n=0}^{\infty} a_n\right) = a_0 + S\left(\sum_{n=1}^{\infty} a_n\right).
\ee
\end{itemize}
Second, we impose the requirement that the distributive property should hold:
\begin{itemize}
\item[(2)] (\textit{Linearity})
\be
S\left(\alpha\sum_{n=0}^{\infty} a_n + \beta\sum_{n=0}^{\infty} b_n\right) = \alpha\, S\left(\sum_{n=0}^{\infty} a_n\right) + \beta\, S\left(\sum_{n=0}^{\infty} b_n\right),
\ee
for $\alpha, \beta\in\mathbb R$. 
\end{itemize}
Now, let us try to apply this generic summation to our favourite divergent series:
\be\bal
s &\equiv S\left( 1-1+1-1+\ldots\right) \\
&= 1 + S\left(-1+1-1+1-\ldots\right) \\
&= 1 - S\left(1-1+1-1+\ldots\right) \\
&= 1-s,
\eal\ee
where finite additivity and linearity have been used in the first and second step, respectively. Again under the hypothesis that $s$ is finite from the start, we deduce $s=1/2$. At this point we see that, since both Euler and Borel summations obey the axioms (1) and (2), they necessarily agree on all the series for which they give finite answer. In particular, they had no choice but to agree on $1-2^p+3^p-4^p+\ldots$ as well!

Let us work out a couple more examples:
\be\bal
s&\equiv S\left(1+2+4+8+\ldots\right) \\
&=1+S\left(2+4+8+\ldots\right)\\
&=1+2S\left(1+2+4+\ldots\right)\\
&=1+2s.
\eal\ee
Hence $s = -1$. Again, consider
\be\bal
s&\equiv S\left(+1-1+0+1-1+0+\ldots\right),\\
s-1&=S\left(-1+0+1-1+0+\ldots\right),\\
s&= S\left(+0+1-1+0+\ldots\right),\\ 
\eal\ee
now, adding up these three equations we find $3s-1=0$, hence $s=1/3$.

We can try applying instead the Euler method to this series: the corresponding power series, which is absolutely convergent for $|x|<1$, can be legitimately rearranged as follows
\be\bal
&1-x+x^3-x^4+x^6-x^7+\ldots\\
=&\left(1+x^3+x^6+\ldots\right) - x\left(1+x^3+x^6+\ldots\right)\\
=&\frac{1}{1-x^3}- \frac{x}{1-x^3} = \frac{1}{1+x+x^2} \xrightarrow[x\to1^-]{} \frac{1}{3},
\eal\ee
again, in agreement with the generic summation method.

 Consider now $1+0-1+0+0+1+0-1+0+0+\ldots$ and let us apply the generic summation tool
\be\bal
s&\equiv S\left(+1+0-1+0+0+1+0-1+0+0+\ldots\right),\\
s-1&=S\left(+0-1+0+0+1+0-1+0+0+\ldots\right),\\
s-1&=S\left(-1+0+0+1+0-1+0+0+\ldots\right),\\
s&=S\left(+0+0+1+0-1+0+0+\ldots\right),\\
s&=S\left(+0+1+0-1+0+0+\ldots\right),\\
\eal\ee
adding up all equations we find $5s-2=0$, hence $s=2/5$. As anticipated, this differs with respect to the sum of $1-1+1-1-\ldots$ indicating a failure of associativity.

We can now try to sum $0-2!+4!-6!+8!-\ldots$ using, for instance, Borel summation:
\be
B\left(\sum_{n=0}^\infty (-1)^n(2n)!\right)= \int_0^\infty e^{-t}\left(\sum_{n=0}^\infty (-1)^n\frac{(2n)!}{n!}t^n\right)dt,
\ee
however the series on the left-hand side does not converge for any positive value of $t$, since 
$n!\sim n^n e^{-n}\sqrt{2\pi n}$ for large $n$ and therefore
\be
\frac{(2n)!}{n!} \sim \sqrt{2}(4n)^ne^{-n}.
\ee
What if we add some zeros here and there? For instance $0!+0-2!+0+4!+0-6!+0+\ldots$ then the Borel sum is instead
\be
\int_0^\infty e^{-t} \left( 1-t^2 + t^4 -t^6+t^8\ldots\right)dt = \int_0^\infty \frac{e^{-t}dt}{1+t^2},
\ee
which is clearly convergent. 

\subsection{Analyticity}
Despite being a more general approach, generic summation does not always work. Consider for instance the following series
\be
s \equiv S\left(1+1+1+\ldots\right) = 1+ S\left(1+1+1+\ldots\right) = 1+s;
\ee
we readily see that this is inconsistent with any finite answer and hence that neither Euler's nor Borel's strategy could be useful to tackle this problem. A common strategy allowing to overcome this difficulty involves defining the Riemann zeta function:
\be
\zeta(s) = \sum_{n=1}^\infty\frac{1}{n^s}\quad \text{ for }s\in\mathbb C \text{ and } \Re (s)>1;
\ee
this function admits an analytic continuation to all $s\in\mathbb C$ such that $\Re(s)\neq1$, and in particular 
\be
1+1+1+\ldots = \zeta(-1) = - \frac{1}{2}.
\ee
The zeta function regularization plays an important role in many areas of theoretical physics: for instance, it allows to give a finite result to the computation of the Casimir energy. Consider two conducting, parallel plates with distance $L$ and surface $A$ (or two boats in the sea, for that matter), and let $x$ be the axis perpendicular to the plates. The electromagnetic field in the vacuum will achieve stationary wave configurations between the plates, \emph{i.e.} the ones with $x$-wavelength given by $\lambda_{x,n}= 2L/n$. The wave vector $\mathbf k$ will be therefore given by two continuous components $k_y$, $k_z$ and the discrete component $k_{x,n}= \pi n/L$.
The total energy stored within the two plates is given by summing over all stationary modes $n=1,2,\ldots$ and by integrating over $k_y$, $k_z$ using the measure given by the continuum limit
\be
dn_y\,dn_z= \frac{A}{(2\pi)^2} dk_y\, dk_z,
\ee
thus, using the dispersion relation $\omega(\mathbf k) = |\mathbf k|$,
\be
E = \sum_{n=1}^\infty \int \frac{A}{(2\pi)^2} dk_y\, dk_z \sqrt{\left(\frac{\pi n}{L}\right)^2+k_y^2+k_z^2}.
\ee
Of course, the integral is UV-divergent but one one can analytically continue it in the following way: consider first 
\be
E^{(s)}=\sum_{n=1}^\infty \int \frac{A}{(2\pi)^2} dk_y\, dk_z \left[\left(\frac{\pi n}{L}\right)^2+k_y^2+k_z^2\right]^{s/2}
\ee 
for $\Re(s)<-2$, so that the integral is convergent, and then consider the limit $s\to1$.
Indeed then we have
\be\bal
E^{(s)}&=\sum_{n=1}^\infty \int \frac{A}{(2\pi)^2} (2\pi) \int_0^\infty \rho\,d\rho \left[\left(\frac{\pi n}{L}\right)^2+\rho^2\right]^{s/2}\\
&=-\frac{A}{2\pi(s+2)}\left(\frac{\pi}{L}\right)^{s+2}\sum_{n=0}^\infty n^{s+2}.
\eal\ee
Finally we regularize this as $s\to1$ using the zeta function
\be
E = - \frac{A\pi^2}{6L^3}\zeta(-3)=-\frac{A\pi^2}{720 L^3}.
\ee
Differentiating $-E$ with respect to the distance $L$ tells us that there is an \emph{attractive} force per unit surface $f_C$, between the parallel plates, given by 
\be
f_C = - \frac{\pi^2}{240 L^4}.
\ee

\subsection{Asymptotic series}
There is a very useful concept which allows one to give a precise meaning to divergent \emph{power} series, namely the notion of being \emph{asymptotic} to some specific function. 
First, let us recall that, given two real functions $f(x)$ and $g(x)$, which admit a limit as $x\to x_0$, we say that $f$ is asymptotic to $g$ as $x\to x_0$, denoted
\be
f(x)\sim g(x),\quad \text{as }x\to x_0,
\ee
whenever
\be
\lim_{x\to x_0} \frac{f(x)}{g(x)} = 1.
\ee
For instance, $\sin x \sim x$ as $x\to 0$, whereas $\Gamma(x+1) \sim x^x e^{-x} \sqrt{2 \pi x}$, as $x\to+\infty$.

Now, consider a power series
\be
\sum_{n=0}^\infty a_n (x-x_0)^n.
\ee
We say that such a series is convergent at a given $x$ if for \emph{fixed} $x$ the sequence of partial sums converges to a finite value:
\be
\lim_{N\to\infty}
\sum_{n=0}^N a_n (x-x_0)^n = f(x) < \infty.
\ee
This defines a function $f(x)$ pointwise in the neighborhood of $x_0$ where the series converges.
As we have seen, in many cases this requirement is too stringent a condition to impose, and we can look for a different concept allowing us to approximate a function with a power series.

Consider now a specific function $f(x)$, defined in a neighborhood of the point $x_0$. We say that the (not necessarily convergent) power series 
\be
\sum_{n=0}^\infty a_n (x-x_0)^n
\ee 
is asymptotic to $f(x)$ as $x\to x_0$, \emph{i.e.}
\be
f(x) \sim \sum_{n=0}^\infty a_n (x-x_0)^n,\quad \text{as }x\to x_0,
\ee
if for fixed $N$
\be
f(x)-\sum_{n=0}^N a_n (x-x_0)^n \sim a_{N+1}(x-x_0)^{N+1},\quad \text{as }x\to x_0,
\ee
or, more explicitly,
\be
\lim_{x\to x_0} \frac{1}{{a_{N+1} (x-x_0)^{N+1}}}\left[f(x)-\sum_{n=0}^N a_n (x-x_0)^n\right] = 1.
\ee
Let us stress that we cannot ask if a given series is ``asymptotic'' without providing any further information: we need to ask whether a series is asymptotic or not to a \emph{given} function (and in a given limit).

A notable example of functions which can be approximated by asymptotic series is that of infinitely differentiable ones. Consider a function $f(x)$ admitting derivatives of arbitrary order at $x_0=0$; then, by repeated use of de l'H\^opital's formula:
	\be
	\lim_{x\to0}\frac{1}{x^{N+1}}\left[f(x)- \sum_{n=0}^N\frac{1}{n!}f^{(n)}(0)x^n\right]=\frac{1}{(N+1)!}f^{(N+1)}(0),	
	\ee
so that
\be
f(x) \sim \sum_{n=0}^{\infty} \frac{1}{n!}f^{(n)}(0) x^n, \quad \text{as }x\to0.
\ee
This also shows that convergent power series, such as the Taylor series, are always asymptotic to the analytic functions they define within their convergence radius.

The relevance of asymptotic series in physical problems is hard to overestimate, since, as we shall see, perturbation series for quantum mechanical systems are almost always divergent, and define instead asymptotic series for the perturbed energy eigenvalues.
\newpage

\section{Continued Functions}
Aside from convergence issues, another very important practical problem we have to face when dealing with perturbative series is the fact that we can only compute a \emph{finite} number of terms $a_n$. For instance, we can only compute a finite number of Feynman diagrams in the perturbative series of quantum field theory. Furthermore, it is typically very hard to guess the general form of the terms in the series expansion, taking inspiration from the first few of them. A strategy to tackle this problem is by employing continued functions, and in particular the theory of continued fraction, which is also at the basis of Pad\'e's summation method.

As a first example showing the convenience of associating a continued function to a given series, consider the continued exponential
\be
a_0 e^{a_1 z e^{a_2 ze^{\ldots}}}.
\ee
We can try to match the expansion of this continued exponential with a power series, order by order in $z$ by formally imposing
\be
a_0 e^{a_1 z e^{a_2 ze^{\ldots}}} = \sum_{n=0}^\infty c_n z^n,
\ee
which yields
\be\bal
c_0 &= a_0,\\
c_1 &= a_1a_0,\\
c_2 &= \frac{1}{2}a_0 a_1^2 + a_0 a_1 a_2,
\eal\ee
and so on.
The advantage of this procedure is that the continued function may well have better convergence properties with respect to the corresponding series. To have a concrete case at hand, one can prove that 
\be\label{eq: cont_exp}
e^{ze^{ze^{\ldots}}} = \sum_{n=0}^\infty \frac{(n+1)^{n-1}}{n!}z^n,
\ee
although the proof is rather hard. Applying the Stirling approximation $n! \sim n^n e^{-n} \sqrt{2\pi n}$ for large $n$, we obtain 
\be
\frac{(n+1)^{n-1}}{n!}z^n \sim \left(1+\frac{1}{n}\right)^n
\!\!
\frac{1}{\sqrt{2\pi n}(n+1)} (ez)^n
\ee
and hence that the series on the right-hand side of \eqref{eq: cont_exp} converges for $|z|< 1/e$ (in fact, it also converges at $|z|=1/e$). However, the continued function on the left-hand side of \eqref{eq: cont_exp} has a much bigger analyticity domain, which also encloses the disk $|z|\le 1/e$. 

\subsection{Continued fractions}
To see how continued fractions are indeed helpful when dealing with a series of which we only know a finite number of terms, we will adopt the following approach. Consider the sequence 
\be
a_0,\ a_1,\ a_2,\ \ldots
\ee and assume for simplicity that $a_n=1$ and $a_n>0$; we will define a procedure to convert $\{a_n\}$ into some $\{b_n\}$, with the advantage that the sequence $\{b_n\}$ will be \emph{easier to guess} from a finite number of terms, compared to the sequence of the $\{a_n\}$!

First, imagine that each $a_n$ could be written as the $2n$th moment of a positive weight function $W(x)$, \emph{i.e.}
\be\label{eq: amoments}
a_n = \int_{-\infty}^{+\infty}x^{2n}W(x)dx,
\ee
where the integrals are assumed to converge.
Second, imagine the $\{b_n\}$ were used to construct polynomials $P_n(x)$ according with the recurrence relation
\be\bal
P_0(x)&=1,\\
P_1(x)&=x,\\
P_{n+1}(x)&= x P_n(x) - b_nP_{n-1}(x).
\eal\ee
As an example, we can calculate the first few $P_n(x)$:
\be\bal
P_2(x)&=x^2-b_1,\\
P_3(x)&=x^3-x(b_1+b_2),\\
P_4(x)&=x^4-x^2(b_1+b_2+b_3)+b_1b_3,
\eal\ee
and so on.
Note that these are all monic polynomials and that $P_n(x)$ is even (odd) whenever $n$ is an even (odd) integer.

Third, we establish a link between the $\{a_n\}$ and the $\{b_n\}$ by requiring that 
$P_n(x)$ should be orthogonal to $P_m(x)$ with respect to the weight function $W(x)$, if $n\neq m$:
\be
\langle P_n, P_m\rangle=\int_{-\infty}^{+\infty}P_n(x) W(x) P_m(x) dx = C_n \delta_{nm},
\ee
for some finite, positive $C_n$. Recalling \eqref{eq: amoments}, this leads to the following constraints:
\be\bal\label{eq: overdetermined_system}
\langle P_0, P_2\rangle&= a_1-b_1=0,\\
\langle P_1, P_3\rangle&= a_2-a_1(b_1+b_2)=0,\\
\langle P_2, P_4\rangle&= a_3 - a_2(2b_1+b_2+b_3)+a_1(b_1+b_2+2b_3)-b_1^2b_3=0, 
\eal\ee
which can be solved recursively to express $a_n$ in terms of $b_n$ (and \textit{vice versa})
\be\bal\label{eq: soluzioni_sovradeterminate}
a_1 &= b_1,\\
a_2 &= b_1 (b_1+b_2),\\
a_3 &= b_1\left[(b_1+b_2)^2+b_2b_3\right]
\eal\ee
and so on.
One might worry that the system \eqref{eq: overdetermined_system} is overdetermined, because the number of orthogonality relations grows like a square, but the number of unknowns grows linearly. For example the orthogonality between $P_0$ and $P_4$ gives another relation involving $b_2$, which has already been determined in \eqref{eq: soluzioni_sovradeterminate}. Nevertheless, let us take a closer look:
\be
\langle P_0, P_4 \rangle= a_2 - a_1(b_1+b_2+b_3)+b_1b_3=a_2-a_1(b_1+b_2) =0.
\ee
Thanks to the cancellation $a_1b_3-b_1b_3=0$, there is no contradiction with \eqref{eq: soluzioni_sovradeterminate}. This illustrates a general phenomenon: The system \eqref{eq: overdetermined_system} admits one and only one recursive solution because all extra equations appearing are in fact redundant and do not give additional independent conditions. 

As we mentioned, given a few $a_n$ it is typically convenient to convert them into the corresponding $b_n$, try to guess the general form of the $b_n$ and then convert back to the starting sequence of $a_n$.

For instance, let us consider the sequence whose first three terms are
\be\label{eq: original_guess}
a_0=1,\ a_1=5,\ a_2=61 .
\ee
It is not at all obvious what the following terms should be. However, converting this into the sequence
\be
b_1=1,\ b_2=4,\ b_3=9
\ee 
allows us to deduce that $b_{n} = n^2$, hence $b_4=16$. Converting back, we find that $a_4=1385$. This is not an obvious guess! We have redescovered the Euler numbers, which are important numbers in discrete mathematics and combinatorics.

In general, one can say that if $a_n$ grows like $n!$, then $b_n$ grows linearly in $n$ and if $a_n$ grows like $(2n)!$, then $b_n$ grows quadratically in $n$ and so on. Thus in the example above, the Euler numbers grow like $(2n)!$. This strategy can also be applied to simplifying the computation of the sequence of Bernoulli numbers.

Up to now, the link between this procedure and continued fractions is not clear. However, let us consider the asymptotic series expansion of the function $f(z)$
\be\label{eq: asymptoticf}
f(z)\sim\sum_{n=0}^\infty (-1)^n a_n z^n,\ \text{as }z\to0
\ee
for $z\in\mathbb C$, where we still assume $a_0=1$ and $a_n>0$.
Consider also the continued fraction
\be
\dfrac{1}{1+\dfrac{b_1 z}{1+\dfrac{b_2 z}{1+\dfrac{b_3 z}{1+\ldots}}}}
\ee
for some $b_n$.
Then, the conditions on $a_n$ and $b_n$ allowing one to match this continued fraction with the series in \eqref{eq: asymptoticf} are exactly the same as the ones given in \eqref{eq: overdetermined_system}, \emph{i.e.} those which define the mapping given at the beginning of this section.

The astonishing property that justifies the importance of this method, is that, even if the radius of convergence of the series in \eqref{eq: asymptoticf} is $|z|=0$, the continued fraction defined by the limit of the sequence
\be
P^0_0=1,\
P^0_1=\dfrac{1}{1+b_1 z},\
P^1_1=\dfrac{1}{1+\dfrac{b_1 z}{1+b_2 z}},\ 
P^1_2=\dfrac{1}{1+\dfrac{b_1 z}{1+\dfrac{b_2 z}{1+b_3 z}}},\ \ldots\ ,
\ee
which is called the main sequence of Pad\'e approximants, 
converges for \textit{any} $z\in\mathbb C$ except for those that lie on the negative real axis.

If the series is also a Stieltjes series, which we shall define and discuss below, the behavior of these approximants is the following. The $P^n_n$ form a decreasing sequence, whereas the $P^n_{n+1}$ form an increasing one. Therefore, both sequences have a limit as $n\to\infty$ and the right answer to the problem whose solution is represented by the original series is trapped between the two limits. Furthermore, when these two limits coincide, and since we have an \emph{alternating} sequence, we can use the Shanks transformation to improve its convergence further!

We can generalize the method of Pad\'e approximants by allowing for more general $P^n_m$. Indeed, we see that $P^1_1$ can be rewritten as the ratio between two polynomials of degree one, and $P^1_2$ as the ratio between a polynomial of degree one and one of degree two. In general, we can match the starting series 
\be
\sum_{n=0}^\infty (-1)^n a_n z^n
\ee
with the Pad\'e approximant given by the ratio of a polynomial of degree $n$ with one of degree $m$
\be
P^n_m(z) = \dfrac{\sum_{j=0}^n b_j z^j}{\sum_{k=0}^m c_k z^k}
\ee
by truncating the series at order $n+m$, multiplying by the denominator of $P^n_m$ and comparing coefficients at each order in $z$. 

As a concluding remark for this section, let us mention that the method of continued fractions can be used not only to sum (\emph{divergent}) perturbation series, possibly even with a finite number of known terms, but also to greatly improve the convergence of Taylor (\emph{convergent}) series: one can be convinced by looking for instance at the power series for
\be
e^z,\quad \frac{1}{\Gamma(z)}\quad \text{or}\quad \frac{\log(1+z)}{z}
\ee
in a neighborhood of the origin. Indeed, the Pad\'e sequence, being the ratio of polynomials, has much more freedom in its behavior, so to speak, and hence manages to approximate the value of these functions with greater accuracy and with a smaller number of terms.

As we shall see in the next section, the convergence properties of the Pad\'e approximants associated to a Stieltjes series are particularly well under control. This is the case for the (asymptotic) series
\be
\sum_{n=0}^\infty (-1)^n n! z^n
\ee
and the Stieltjes property also allows us to sum the perturbation series for the ground-state energy of the one-dimensional anharmonic oscillator, defined by the Hamiltonian 
\be
H = \frac{p^2}{2} + \frac{x^2}{2} + \eps {x^4},
\ee
whose first few terms are
\be
E_0(\epsilon)=\frac{1}{2}+\frac{3}{4}\epsilon - \frac{21}{8}\epsilon^2+\frac{333}{16}\epsilon^3-\ldots\ \ .
\ee
It is an instructive exercise to compute this perturbative expansion, since it corresponds to that of a one-dimensional $\varphi^4$ model. In the (Euclidean) path integral approach, the zero-point energy is given by the expression
\be
e^{-E_0 T} = \int \mathcal D x\, \exp\left\{-\int \left(\frac{\dot x^2}{2}+\frac{x^2}{2}+\eps x^4\right)dt\right\}.
\ee
For such a theory, the free propagator is the solution to $-\ddot G(t) + G(t)=\delta(t)$, \emph{i.e.} 
\be
\omega^2 \tilde G(\omega)+ \tilde G(\omega) = 1;
\ee
in Fourier space, which gives
\be
G(t) = \frac{1}{2\pi}\int_{-\infty}^{+\infty} \frac{e^{i\omega t}}{1+\omega^2}\,d\omega = \frac{e^{-|t|}}{2}.
\ee
The interaction vertex is $-4! \eps$.

Now we can calculate 
\be\bal
E_0(\eps) &= -\frac{1}{T}\log\left(\text{vacuum bubbles}\right)\\
&= - \frac{1}{T} \left(\text{connected vacuum bubbles}\right).
\eal\ee
To zeroth order (\emph{i.e.} at the unperturbed level) we have
\be
E_0^{(0)} = \frac{1}{2}.
\ee

\begin{fmffile}{diagram}

The first-order correction is given by the diagram with only one interaction vertex:
\be
\begin{gathered}
\begin{fmfgraph*}(100,70)
\fmftop{l,r}
\fmfbottom{s,t}
\fmf{phantom}{l,v,r}
\fmf{phantom}{s,v,t}
\fmffreeze
\fmf{plain,right=90}{v,v}
\fmf{plain}{v,v}
\end{fmfgraph*}
\end{gathered}
E_0^{(1)} =-\frac{1}{T}(-4!\eps) \frac{1}{8}\int_{-T/2}^{T/2}G(0) G(0) dt = \frac{3}{4}\eps,
\ee
where $1/8$ takes into account the symmetries of the diagram ($1/2$ for each flipping of the tadpole lines, and again $1/2$ for the reflection symmetry around the horizontal axis).

To second order we have two connected diagrams, yielding respectively
\vspace{-15pt}$$
\begin{gathered}
\begin{fmfgraph*}(100,70)
\fmfleft{l,m}
\fmfright{r,s}
\fmftop{t0,t1,t2,t3}
\fmfbottom{b0,b1,b2,b3}
\fmf{phantom}{t1,v1,b1}
\fmf{phantom}{t2,v2,b2}
\fmffreeze
\fmf{plain,right}{v1,v2,v1}
\fmf{plain,tension=0.8,right=270}{v1,v1}
\fmf{plain,tension=0.8,right}{v2,v2}
\end{fmfgraph*}
\end{gathered}
-\frac{1}{T} (-4!\eps)^2 \frac{1}{16} \int_{-T/2}^{T/2}\int_{-T/2}^{T/2} G(t_1-t_2) G(t_2-t_1)G(0)^2 dt_1 dt_2 = - \frac{9}{4}\eps^2,
\vspace{-15pt}$$
and
\vspace{15pt}$$
\begin{gathered}
\begin{fmfgraph*}(75,55)
\fmfleft{l,m}
\fmfright{r,s}
\fmftop{t1,t2}
\fmfbottom{b1,b2}
\fmf{phantom}{t1,v1,b1}
\fmf{phantom}{t2,v2,b2}
\fmffreeze
\fmf{plain,right}{v1,v2,v1}
\fmf{plain, right=.5}{v1,v2}
\fmf{plain, left=.5}{v1,v2}
\end{fmfgraph*}
\end{gathered}\ \ \
-\frac{1}{T} (-4!\eps)^2 \frac{1}{2} \int_{-T/2}^{T/2}\int_{-T/2}^{T/2} G(t_1-t_2)^2 G(t_2-t_1)^2 dt_1 dt_2 = -\frac{3}{8}\eps^2.
\vspace{5pt}
$$
Hence,
\be
E_0^{(2)}=-\frac{21}{8}\eps^2.
\ee
To third order we have four diagrams:
$$
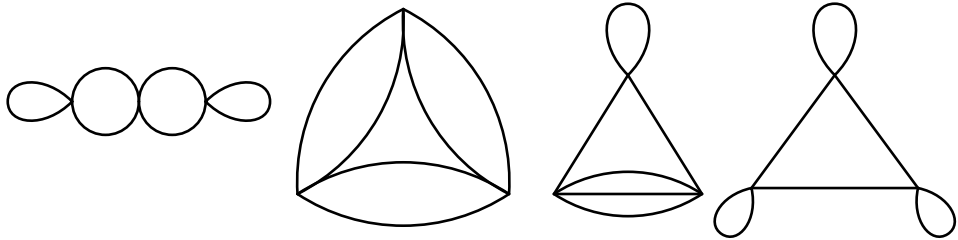
\begin{fmfgraph*}(100,70)
\fmfleft{l,m}
\fmfright{r,s}
\fmftop{t0,t1,t,t2,t3}
\fmfbottom{b0,b1,b,b2,b3}
\fmf{phantom}{t1,v1,b1}
\fmf{phantom}{t,v,b}
\fmf{phantom}{t2,v2,b2}
\fmffreeze
\fmf{plain,right}{v1,v,v1}
\fmf{plain,right}{v,v2,v}
\fmf{plain,tension=0.8,right=270}{v1,v1}
\fmf{plain,tension=0.8,right}{v2,v2}
\end{fmfgraph*}
\begin{fmfgraph*}(100,70)
\fmftop{a}
\fmfleft{c,d}
\fmfright{b,d1}
\fmf{plain, left=.3}{a,b}
\fmf{plain, left=.3}{b,c}
\fmf{plain, left=.3}{c,a}
\fmf{plain, right=.3}{a,b}
\fmf{plain, right=.3}{b,c}
\fmf{plain, right=.3}{c,a}
\end{fmfgraph*}
\begin{fmfgraph*}(70,45)
\fmfleft{ld,lu}
\fmfright{rd,ru}
\fmf{phantom}{ru,v,lu}
\fmffreeze
\fmf{plain}{ld,v,rd,ld}
\fmf{plain}{v,v}
\fmf{plain,left=.3}{ld,rd}
\fmf{plain,right=.3}{ld,rd}
\end{fmfgraph*}
\ \ \ 
\begin{fmfgraph*}(70,45)
\fmfleft{ld,lu}
\fmfright{rd,ru}
\fmfbottom{bl,br}
\fmf{phantom}{lu,a,ru}
\fmf{phantom}{rd,b,br}
\fmf{phantom}{bl,c,ld}
\fmffreeze
\fmf{plain}{a,b,c,a}
\fmf{plain}{a,a}
\fmf{plain}{b,b}
\fmf{plain}{c,c}
\end{fmfgraph*}
$$
whose evaluation is given by
\be\bal
&-\frac{1}{T} (-4!\eps)^3 \frac{1}{32} \int\int\int G(t_1-t_2)G(t_2-t_3)G(t_3-t_2)G(t_2-t_1)G(0)^2dt_1 dt_2 dt_3,\\
&-\frac{1}{T} (-4!\eps)^3 \frac{1}{48} \int\int\int G(t_1-t_2)^2 G(t_2-t_3)^2 G(t_3-t_1)^2 dt_1 dt_2 dt_3,\\
&-\frac{1}{T} (-4!\eps)^3 \frac{1}{24} \int\int\int G(t_1-t_2)^3 G(t_2-t_3) G(t_3-t_1) G(0)  dt_1 dt_2 dt_3,\\
&-\frac{1}{T} (-4!\eps)^3 \frac{1}{48} \int\int\int G(t_1-t_2) G(t_2-t_3) G(t_3-t_1) G(0)^3 dt_1 dt_2 dt_3,
\eal\ee
giving 
\be
E_{0}^{(3)}=\frac{333}{16}\eps^3.
\ee
Remarkably enough, there exists a formula for the asymptotic behavior of the coefficients in this expansion, namely
\be\label{eq: proviso}
a_n \sim \frac{(-1)^{n+1}\sqrt 6}{\pi^{3/2}} 3^n \Gamma\left(n+\frac{1}{2}\right).
\ee
In the next section we will sketch the idea of the proof of this formula.

\end{fmffile}
As a final comment, let us mention that, for the $\eps x^4$ perturbation series, the number $N(n)$ of \emph{all} diagrams occurring at order $n$ (counted with their multiplicity due to symmetry) is given by a simple formula. Let us consider the set of all $n$ vertices, which in total have $4n$ legs, and let us first pretend that we are able to distinguish each vertex and leg; picking one of these legs (it does not matter which one) we will have $4n-1$ possible other legs to which it can be connected. After this choice, the next leg we choose to connect will have $4n-3$ possibilities and so on giving in total $(4n-1)(4n-3)\cdots 3 \cdot 1 = (4n-1)!!$ Now we must recall that in fact vertices and legs are indistinguishable, and we can account for this fact by dividing by the number of their permutations: $(4!)^n$ for the legs and $n!$ for the vertices.
To sum up,
\be
N(n) = \frac{(4n-1)!!}{n!(4!)^n}. 
\ee

Another way of getting to the same result is using the functional integral representation for the zero-dimensional $\varphi^4$ model. The normalized coefficient of $\eps^n$ in the integral
\be
I\equiv\int_{-\infty}^{+\infty} e^{-x^2/2+\eps\, x^4/4!} dx
\ee 
will give the desired number;
this reduces to computing Gaussian integrals
\be
I = \sum_{n=0}^\infty \frac{\eps^n}{n!(4!)^n} \int_{-\infty}^{+\infty} x^{4n}e^{-x^2/2}dx =  
\sum_{n=0}^\infty \frac{\eps^n}{n!(4!)^n}
(-2)^{2n}\left(\frac{d}{d\alpha}\right)^{2n}\int_{-\infty}^{+\infty}e^{-\alpha x^2/2}dx\bigg|_{\alpha=1}.
\ee

\subsection{Herglotz functions}
Before diving into the properties of physically relevant perturbation series, such as those defining the energy levels for certain quantum mechanical systems, we need to recall the definition of a Herglotz function.

A complex function $f(z)$ is \textit{Herglotz} if its imaginary part has the same sign as the imaginary part of $z$: more precisely
\be\begin{aligned}
\text{Im} f(z) > 0\quad  &(z>0),\\
\text{Im}f(z) = 0 \quad &(z=0),\\
\text{Im}f(z) < 0\quad &(z<0).
\end{aligned}
\ee
The Herglotz property arises naturally in quantum-mechanical perturbation problems: consider the perturbed Hamiltonian
\be\label{eq: Hamiltonian}
H(\epsilon)=\frac{p^2}{2} + V(x) + \epsilon W(x)
\ee 
for a small complex parameter $\epsilon$, where $V(x)$ is a real potential and $W(x)$ is a real perturbation, which can be taken positive; the associated eigenvalue problem reads
\be
\left[\frac{p^2}{2} + V(x) + \epsilon W(x)\right] |\psi\rangle = E(\epsilon) |\psi\rangle;
\ee
applying now $\langle \psi|$ from the left, and taking the imaginary part of both sides, 
\be
\text{Im}\, E(\epsilon) = \langle \psi|W(x)|\psi\rangle\, \text{Im}\,\epsilon.
\ee
This shows that, in perturbation theory, $E(\eps)$ \emph{is} Herglotz.

Another example of Herglotz functions is that of any function defined by
\be
f(s) = - \int_0^\infty \frac{W(t)}{1+st}dt,
\ee
where $W(t)$ is a positive weight function, for any non-negative $s\in\mathbb C$: indeed, multiplying and dividing by $1+\bar s t$,
\be
f(s) = -\int_0^\infty \frac{W(t)}{|1+st|^2}dt - \bar s \int_0^\infty \frac{W(t)t}{|1+st|^2}dt,
\ee
so clearly Im$f(s)$ and Im$s$ have the same sign.

A remarkable property of Herglotz functions is that any \emph{entire} (\emph{i.e.} everywhere-analytic) Herglotz function $f(z)$ is at most linear. To see this, consider such an $f(z)$ and expand it into its Taylor series around the origin
\be
f(z) = a_0 + \sum_{n=1}^\infty a_n z^n.
\ee
First of all, by the Herglotz property, $f(z)$ must be real when $z$ is real, and hence all the $a_n$ must be real, for $n=0,1,2,\ldots$ (indeed, they are given by the derivatives of a real function). 
Now, let us expand $z=\rho\, e^{i\theta}$ in polar form 
\be
f(z) =a_0 + \sum_{n=1}^\infty a_n \rho^n e^{in\theta}
\ee
and take the imaginary part of the previous equation:
\be\label{eq: expansion_f}
\text{Im}f(z)=\sum_{n=1}^\infty a_n \rho^n \sin n\theta.
\ee
Now, notice that $m \sin \theta \pm \sin m\theta$, for $m$ integer and greater than one, has the same sign\footnote{
To prove this fact, it suffices to show that 
\be
(m\sin\theta \pm \sin m\theta)\sin\theta = m \sin^2 \theta \left( 1\pm\frac{\sin m\theta}{m \sin\theta}\right)
\ee	
is always positive (or zero);
but indeed, using the Euler formulas,
\be
\left|\frac{\sin m\theta}{m \sin\theta}\right|= \frac{1}{m} \left| \frac{1-e^{-i2m\theta}}{1-e^{-i2\theta}}\right|=\frac{1}{m}\left|\sum_{k=0}^{m-1}e^{-i2k\theta}\right| \le \frac{1}{m}\cdot m=1.
\ee
	} 
as $\sin\theta$: therefore, by the Herglotz property
\be\label{eq: confronto}
\left(m \sin \theta \pm \sin m\theta\right) \text{Im} f(z) \quad\text{and}\quad \sin\theta\, \text{Im}\,z
\ee
also have the same sign. By the expansion \eqref{eq: expansion_f}, the first quantity in \eqref{eq: confronto}, integrated from $0$ to $\pi$ in $d\theta$, reads
\be
\int_0^\pi\sum_{n=1}^\infty a_n \rho^n\left(m \sin \theta \pm \sin m\theta\right) \sin n\theta d\theta = \frac{\pi}{2}\left(a_1 \rho \pm a_m \rho^m\right),
\ee
whereas the second quantity in \eqref{eq: confronto} gives just
\be
\rho \int_0^\pi \sin^2\theta d\theta= \rho \frac{\pi}{2},
\ee
that is, a positive quantity.
This clearly leads to a contradiction, since the $\pm a_m \rho^m$ can attain arbitrarily large positive or negative values, unless $a_m=0$ for $m$ greater than one.

This observation enables us to give an interpretation to the occurrence of singularities in quantum-mechanical perturbation theory. Considering the Hamiltonian in \eqref{eq: Hamiltonian}, we have shown that the energy levels $E(\eps)$ are Herglotz functions of the perturbation parameter $\eps$. Now, in perturbation theory, we usually expand $E$ as 
\be\label{eq: generale}
E(\eps)=\sum_{n=0}^\infty a_n \eps^n,
\ee
and assuming that this series converges for all $\eps$ is equivalent to requiring $E(\eps)$ to be an entire analytic function. By our simple theorem on the properties of Herglotz functions, we know that this situation can only occur in the trivial case where $E(\eps)$ is linear!

Therefore, we typically expect our quantum mechanical perturbation series to diverge, at least for some $\eps$; in fact, such singularities frequently occur at the origin, and the series \eqref{eq: generale} is only asymptotic to the function $E(\eps)$ as $\eps\to0$. 

\subsection{Singularities in perturbation theory}

Another useful example allowing us to look at the issue of singularities in the perturbation series from a different perspective is the following: consider the two-dimensional Hamiltonian
\be
H = \left(
\begin{matrix}
	a & \eps c\\
	\eps c & b
\end{matrix}
\right)
\ee 
where $a$, $b$ and $c$ are real numbers, and $\eps$ is a complex perturbation parameter. The energy eigenvalues associated to this are given by the associated secular equation:
\be
\text{det}\left(
\begin{matrix}
	a-E & \eps c\\
	\eps c & b-E
\end{matrix}
\right)=0
\ee
\emph{i.e.}
\be\label{5.56}
E_{\pm}(\eps)=\frac{a+b \pm \sqrt{(a-b)^2+4\eps^2c^2}}{2}.
\ee
As we expected, this expression has branch-point singularities in $\eps$ at
\be
\pm i \frac{a-b}{2c}, 
\ee 
whose branch cuts can be deformed to a single cut in the complex plane, joining these two points. 

Now, since these branch cuts originate form the presence of a square root, an alternative picture of this expression can be given replacing the complex plane with a two-sheeted Riemann surface: crossing the cut once will correspond to moving to a new $\mathbb C$-sheet, whereas crossing it twice means going back to the original one. Thus, we can write
\be\label{5.58}
E(\eps)=\frac{a+b + \sqrt{(a-b)^2+4\eps^2c^2}}{2},
\ee
where $E_\pm(\eps)$ in \eqref{5.56} are obtained by evaluating the expression in \eqref{5.58} on the first or second sheet. In particular, the unperturbed eigenvalues, $a$ and $b$, are given by $E_\pm(0)$ so that the quantization of the original system can be interpreted as a \emph{consequence} of the existence of multiple Riemann sheets.

This phenomenon is quite general, and indeed this reasoning can be extended to $n\times n$ matrix Hamiltonians or harmonic oscillators. In the latter case, for instance, \emph{i.e.} for the Hamiltonian
\be
H = \frac{p^2}{2} + \frac{x^2}{2} + \eps {x^4},
\ee
one can show that there exist countably many branch cut singularities: crossing them again leads from one sheet to another and hence from the $n$th to the $m$th energy level.

\subsection{Stieltjes series and functions}
A series is a \textit{Stieltjes series} if it has the form
\be
\sum_{n=0}^\infty (-1)^n a_n x^n,
\ee
where $a_n$ (assumed to be finite) is the $n$th moment of a positive weight function $W(t)$
\be
a_n = \int_0^\infty W(t) t^n dt.
\ee
Examples of this kind of series are, for instance,
\be
\sum_{n=0}^\infty (-1)^n n! x^n,
\ee
since
\be
n! = \int_0^\infty e^{-t}t^{n} dt,
\ee
or also
\be
\frac{\log(1+x)}{x} = \sum_{n=0}^\infty (-1)^n \frac{x^n}{n},
\ee
where
\be
\frac{1}{n}= \int_0^\infty \chi_{[0,1]}(t)t^n dt
\ee
and $\chi_{[0,1]}(t)$ is the characteristic function of $[0,1]$.

As we have mentioned, a Stieltjes series has the property that the associated main sequence of Pad\'e approximants splits into a decreasing and an increasing sequence. Furthermore, these two subsequences converge to the \emph{same} limit when
$a_n$ grows no faster than $(2n)!\, c^n$, where $c$ is a constant.
This property is known as  \textit{Carleman's condition}.

We say that $f(z)$ is a Stieltjes function if it has the form 
\be\label{eq: Stieltjes_f}
f(z) = \int_0^\infty \frac{W(t)}{1+zt} dt,
\ee
where again $W(t)$ is a positive weight function with finite moments. Stieltjes functions $f(z)$ exhibit the following four properties in the complex plane, except on the negative real axis (that is, in the \emph{cut plane}):
\begin{enumerate}
	\item $f(z)$ is analytic; 
	\item $f(z)\to0$ as $z\to\infty$;\footnote{This requirement can be relaxed to the condition that $f(z)$ grows like $z^a$ as $z\to\infty$, but then it is the subtracted function that is a Stieltjes function. For instance if $0<a<1$, then $[f(z)-f(0)]/z$ is a function of Stieltjes.}
	\item $f(z)$ has an asymptotic series expansion
	\be
	f(z) \sim \sum_{n=0}^\infty(-1)^n a_n z^n,\quad \text{as }z\to0,
	\ee
	where $a_n>0$;
	\item $-f(z)$ is Herglotz.
\end{enumerate}
Note that property 1 follows from direct computation of the derivatives of \eqref{eq: Stieltjes_f}, for instance
\be
f'(z) = - \int_0^\infty \frac{W(t)t}{(1+zt)^2}dt,
\ee
property 3 follows from the existence of such derivatives (and the fact that their sign is alternating),
property 2 is clear from \eqref{eq: Stieltjes_f} itself and property 4 follows by multiplying and dividing the integrand by $1+\bar z t$.

The converse is also true: \emph{any} function $F(z)$ satisfying conditions $1$--- $4$ above is a Stieltjes function. To see this, first, by property 1, we can use Cauchy's integral representation to rewrite F(z) as
\be
F(z) = \frac{1}{i2\pi}\oint_\Gamma \frac{F(\zeta)}{\zeta-z}d\zeta,
\ee
for any $z$ in the cut plane and any contour enclosing $z$ which does not cross the negative real axis.

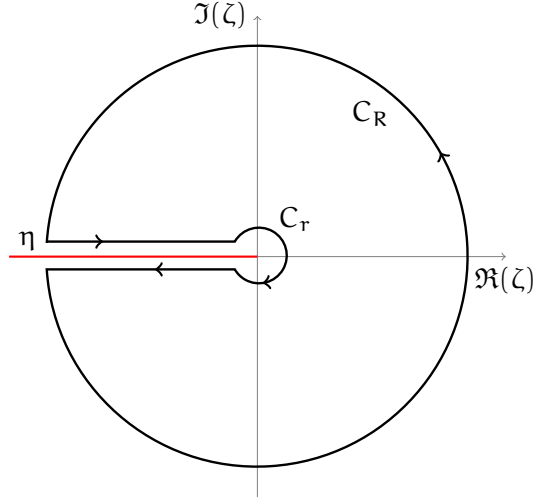
\begin{figure}
	\begin{center}
\begin{tikzpicture}
[decoration={markings,
	mark=at position 0.75cm with {\arrow[line width=1pt]{>}},
	mark=at position 4cm with
	{\arrow[line width=1pt]{>}},
	mark=at position 5.5cm with
    {\arrow[line width=1pt]{>}},
	mark=at position 17cm with {\arrow[line width=1pt]{>}}
}
]
\draw[help lines,->] (-3.3,0) -- (3.3,0) coordinate (xaxis);
\draw[help lines,->] (0,-3.2) -- (0,3.2) coordinate (yaxis);
\draw[red, thick] (0,0) -- (-3.3,0);

\path[draw,line width=0.9pt,postaction=decorate] (-2.8,0.2) node [left] {$\eta$} -- 
+ (2.5,0) 
arc(150:-150:.37) -- 
+ (-2.5,0)
arc(-176.4:176.4:2.8);

\node[below] at (xaxis) {$\Re(\zeta)$};
\node[left] at (yaxis) {$\Im(\zeta)$};
\node at (.5,.5) {$C_{r}$};
\node at (1.5,1.9) {$C_{R}$};
\end{tikzpicture}
\end{center}
\caption{The keyhole contour $\Gamma$ in the cut plane $\mathbb C \setminus \{z<0\}$.}
\label{Keyh}
\end{figure}

 In particular, we can choose $\Gamma$ to be a keyhole contour whose inner circle $C_r$ and outer circle $C_R$ have radii $r$ and $R$ respectively, and with separation $\eta$ from the negative real axis (see Fig. \ref{Keyh}). By property 2, this gives
\be
F(z) = \frac{1}{i2\pi} \int_{-\infty}^{0} \frac{F(t+i\eta)-F(t-i\eta)}{t-z} dt,
\ee
as $R\to\infty$ and $r\to 0$.
Letting $\eta\to0$ we have
\be
F(z) = \frac{1}{\pi} \int_{-\infty}^{0} \frac{D(t)}{t-z}dt,
\ee
where
\be
D(t)\equiv \frac{1}{2i}\lim_{\eta\to0^+}\left[F(t+i\eta)-F(t-i\eta)\right].
\ee 
Now, let us rewrite $F(z)$ as 
\be
F(z) = \frac{1}{\pi} \int_{-\infty}^{0} \frac{D(t)}{t}\frac{1}{1-z/t}dt
=\sum_{n=0}^\infty\int_{-\infty}^{0}\frac{D(t)}{\pi t^{n+1}}dt\, z^n.
\ee
Notice that all integrals appearing in this equation are convergent, since they give the asymptotic expansion coefficients of property 3, and that $D(t)$ is positive by the Herglotz property 4.

These considerations have very interesting physical applications: indeed, consider for instance the perturbation series for the ground-state energy of the anharmonic oscillator
\be
E_0(\epsilon)=\frac{1}{2}+\frac{3}{4}\epsilon - \frac{21}{8}\epsilon^2+\frac{333}{16}\epsilon^3-\ldots
\ee
and define
\be
F(\eps) = \frac{E_0(\eps)-E_0(0)}{\eps} = \frac{3}{4} - \frac{21}{8}\eps+ \frac{333}{16}\eps^2-\ldots\ \ .
\ee
One can show that $F(\eps)$ satisfies all properties 1--- 4 (this subtraction trick is needed in particular to ensure property 3) and hence one obtains the following integral representation for the expansion coefficients:
\be
a_n = \int_{-\infty}^{0}\frac{D(t)}{\pi t^{n+1}}dt,
\ee
which can be interpreted as a dispersion relation.
\renewcommand{\arraystretch}{1.5}
\begin{table}
	\begin{center}
		\begin{tabular}{ |l| c|| l| c | }
			\hline
			$P^0_1$ & 0.66667   & $P^1_1$ & 0.95600  \\
			\hline 
			$P^1_2$ & 0.73385  & $P^2_2$ & 0.87411  \\
			\hline
			$P^2_3$ & 0.76506   & $P^3_3$ &  0.84110  \\
			\hline
			$P^3_4$ & 0.78102  & $P^4_4$ & 0.82529\\
			\hline
		\end{tabular}
		\caption{Pad\'e approximants for the ground-state energy of the anharmonic oscillator.}
		\label{table_anh}
	\end{center}
\end{table}

A good approximation for $D(t)$ can be obtained via Wentzel-Kramers-Brillouin (WKB) methods for the tunneling effect (recall that $t<0$ means negative coupling), which yields \eqref{eq: proviso}. The important fact to notice is that 
$a_n$ behaves like
\be
\quad n!\, 3^n\quad  (n\to\infty),
\ee
which satisfies Carleman's condition. Hence, we can conclude that the perturbation series of the anharmonic oscillator has a convergent associated Pad\'e sequence as illustrated by Table \ref{table_anh}.

This result on the large-order behavior of one-dimensional perturbation theory can be also obtained by studying Feynman diagrams with many vertices in terms of a suitable continuum limit (\emph{viz.}
C. M. Bender and T. T. Wu, \textit{Phys. Rev. Lett.} \textbf{37}, 117 (1976)).
\end{document}